\documentclass[12pt, preprint]{aastex62}

\newcommand{\reff}{\,$R_{eff}$}
\newcommand{\fa}{$f_A$}
\newcommand{\Mr}{\,$M_r$}   

\newcommand{\msun}{\hbox{${M}_{\odot}$}}

\newcommand{\be}{\begin{equation}}
\newcommand{\ee}{\end{equation}}
\newcommand{\ba}{\begin{eqnarray}}
\newcommand{\ea}{\end{eqnarray}}

\newcommand{\simgt}{\lower 2pt \hbox{$\, \buildrel {\scriptstyle >}\over {\scriptstyle\sim}\,$}}
\newcommand{\simlt}{\lower 2pt \hbox{$\, \buildrel {\scriptstyle <}\over {\scriptstyle\sim}\,$}}
\newcommand{\ls}{\lower 2pt \hbox{$\;\scriptscriptstyle \buildrel<\over\sim\;$}}
\newcommand{\gs}{\lower 2pt \hbox{$\;\scriptscriptstyle \buildrel>\over\sim\;$}}

\newcommand{\comment}[1]{}  

\graphicspath{{./}{figures/}}

\submitjournal{ApJL}

\shorttitle{AGN Abundance in Cosmic Voids}
\shortauthors{Mishra et al.}

\begin{document}

\title{Active Galactic Nuclei Abundance in Cosmic Voids}

\email{hora.mishra@ou.edu}

\author[0000-0002-6821-5927]{Hora D. Mishra}
\affil{Homer L.\ Dodge Department of Physics and Astronomy,
University of Oklahoma, Norman, OK 73019, USA}

\author[0000-0001-9203-2808]{Xinyu Dai}
\affil{Homer L.\ Dodge Department of Physics and Astronomy,
University of Oklahoma, Norman, OK 73019, USA}
\email{xdai@ou.edu}

\author{Eduardo Guerras}
\affil{Homer L.\ Dodge Department of Physics and Astronomy,
University of Oklahoma, Norman, OK 73019, USA}



\begin{abstract}

The abundance of active galactic nuclei (AGN) in cosmic voids is relatively unexplored in the literature, but can potentially provide new constraints on the environmental dependence of AGN activity and the AGN-host co-evolution.  We investigated AGN fraction in one of the largest samples of optically selected cosmic voids from SDSS Data Release 12 for redshift range $0.2-0.7$ for moderately bright and bright AGN. We separated inner and outer void regions based on the void size, given by its effective void radius. We classified galaxies at a distance $<0.6 R_{eff}$ as inner void members and galaxies in the interval \textbf{$0.6<R/R_{eff}<1.3$ }as outer void galaxies. We found higher average fractions in the inner voids (4.9$\pm$0.7)\% than for their outer counterparts (3.1$\pm$0.1)\% at $z>0.42$, which clearly indicates an environmental dependence. This conclusion was confirmed upon further separating the data in narrower void-centric distance bins and measured a significant decrease in AGN activity from inner to outer voids for $z>0.42$. At low redshifts ($z<0.42$), we find very weak dependence on the environment for the inner and outer regions for two out of three bins. We argue that the higher fraction in low-density regions close to void centers relative to their outer counterparts observed in the two higher redshift bins suggests that there may occur more efficient galaxy interactions at one-to-one level in voids which may be suppressed in denser environments due to higher velocity dispersions. It could also indicate less prominent ram pressure stripping in voids or some intrinsic host or void environment properties.

\end{abstract}



\section{Introduction} \label{sec:intro}
Galaxies are found in different large-scales structures that form the cosmic web of the Universe: the overdense regions called galaxy clusters which expand out into filamentary structures, which in turn form boundaries of vast underdense regions, known as the cosmic voids (\citealt{bond1996}; \citealt{pan2012}), mostly devoid of matter (dark matter and baryons).
Voids have a radius of tens of Mpc to over one hundred and constitute the majority of the universe's volume (e.g. \citealt{zeldovich1982}; \citealt{pisani2019}). They contain few, isolated galaxies and not much gas. Due to their underdense nature, void properties are dominated by unclustered components, such as dark energy. 
In voids, there is a spherical outward motion of matter toward the walls and filaments (e.g. \citealt{hamaus2016}).

Most, if not all, galaxies have a supermassive blackhole (SMBH) at their center. Of these, a fraction of galaxies are active where their central engine is accreting mass from the dense central region of the galaxy at a sufficiently high rate. It is not well understood why only a few SMBHs show such high nuclear activity. 
There is strong evidence that galaxy evolution is closely related to its environment, influencing its star formation, morphology etc. (e.g. \citealt{marconi2003}; \citealt{einasto2008}; \citealt{2016ApJ...825...72A} ).
Possible triggering mechanism(s), such as major and minor mergers (e.g. \citealt{2005Natur.433..604D}; \citealt{2015MNRAS.451.2968F}), tidal effects (\citealt{1996Natur.379..613M}), disc instability (\citealt{2009ApJ...703..785D}), may be significant in driving AGN activity by supplying cold gas to the central black hole, thus triggering it. 
In the AGN phase model, where most black holes in galaxies undergo an intense activity period, the lifetimes of emission at AGN luminosities are estimated to be in the range $10^6-10^8$ years. Based on the models of black hole growth via gas inflows, the strong accretion phase lasts for $\sim 10^8$ years (\citealt{2000MNRAS.311..576K}; \citealt{2005ApJ...625L..71H}; \citealt{2009ApJ...690...20S}).  
Understanding galaxy evolution requires the study of complex external and internal processes, such as mergers, gas stripping, disc instability, that might have strong influence on its evolution and AGN activity. 

The role of the galaxy's environment in AGN triggering is widely debated. In galaxy clusters and filaments, a combination of factors, such as galaxy rich environment which leads to more galaxy interactions and mergers, extreme conditions in the cluster's gravitational potential well, concentration of cold gas in the cluster halo, and Ram-pressure stripping from the ICM, determines AGN activity (\citealt{1972ApJ...176....1G}; \citealt{1980ApJ...237..692L}; \citealt{2005Natur.433..604D}). 
Many studies such as \citet{2019MNRAS.487.2491E} and \citet{mishra2020} find evidence for variation in fraction of AGN with high and low-density environments, while several others see no to very weak correlation between the two (e.g. \citealt{2003ApJ...597..142M}; \citealt{2013MNRAS.429.1827P}; \citealt{2019MNRAS.tmp.1665M}).
The influence of environmental factors on AGN fraction is tied to the evolution of galaxies and clusters over cosmic time. Several studies show that cluster environments have been dynamically evolving over the history of the universe and that has strong influence on AGN fractions in both clusters and fields (\citealt{2007ApJ...664L...9E}; \citealt{2017MNRAS.465.2531B}). In the local universe, there is found to be evidence for anti-correlation between the AGN fraction and galaxy density (e.g. \citealt{2017MNRAS.472..409L}; \citealt{mishra2020}). Other studies have found comparable AGN fractions in clusters and fields for low-luminosity quasars (\citealt{2010ApJ...723.1447H}). However, at higher redshifts, the AGN evolution is seen to follow a different evolutionary path (\citealt{2013ApJ...768....1M}). Cosmic conditions at higher redshifts, such as galaxy and cluster morphologies, presence of denser ICM, dominance of dark matter, have greatly impacted the large-scale matter distribution.
Studies show that the clusters have undergone significant evolution over the history of the universe which has influenced AGN abundance (e.g. \citealt{2007ApJ...664L...9E}; \citealt{2017MNRAS.465.2531B}; \citealt{mishra2020}). 

In contrast to clusters, void dynamics have remained largely unchanged. In void environment, galaxy evolution is more likely to be dominated by secular (in-situ) processes, because mergers and gas stripping don't occur as much due to the very low galaxy density (\citealt{porqueres2018}; \citealt{habouzit2020}). This would imply less evolved galaxies are contained in the voids, which must have formed at later times and followed different evolutionary path from cluster and group galaxies. Thus, voids form an ideal environment to investigate galaxy evolution and its dependence on secular processes in the absence of external processes, like mergers, which dominate in the high-density environments. Studies have shown that voids comprise of fainter, bluer galaxies, with higher star formation rates, than their overdense counterparts (e.g. \citealt{hoyle2005}; \citealt{goldberg2005}; \citealt{bruton2020}). 
Previous works investigating AGN abundance and accretion activity in voids have also found mixed results that depend on host properties. For example, \citet{constantin2008} found that AGN are more common in voids than walls for moderately luminous and massive galaxies (\Mr $\sim$ --20, log M/\msun $<$ 10.5), but the AGN abundance is comparable for brighter hosts (\Mr $< -20$). \citet{kauffmann2003} find a decreasing fraction of strong AGN in massive galaxies as a function of density, which provides evidence for an environmental dependence.  

Compared to the rich literature available for studying AGN activity in galaxy clusters, groups, and fields, there have not been many studies that have looked at the AGN abundance and AGN evolution in cosmic voids. Our motivation behind this study is to investigate the presence of environmental influence, if any, on the triggering of the supermassive black hole in the absence of ``nurture" processes, and contribute to the understanding of AGN in the most underdense regions of the universe. We look at one of the largest spectroscopic void samples to study the fraction of optical AGN in voids.  This is crucial to understand how the different local and global environmental processes found in baryon-devoid voids may trigger or suppress AGN activity. 


\section{DATA AND METHODOLOGY} \label{sec:sec2}
We investigate AGN fraction using 1,228 cosmic voids, 166,067 member galaxies (non-AGN) and 3,100 AGN in the voids. 
In the following subsections, we describe the data and the method.
Based on the dataset used in this study, the analysis of this paper is limited to investigate the AGN fraction at the luminous end (\Mr $\leq -22$) in inner and outer regions of voids in the local-intermediate universe ($0.2<z<0.7$).

\subsection{Galaxy Catalogs}
The galaxies were selected from the Baryon Oscillation Spectroscopic Survey (BOSS; \citealt{dawson2013}) which is part of the Sloan Digital Sky Survey III (SDSS III; \citealt{eisenstein2011}). SDSS is a multi-band imaging and spectroscopic survey that uses a 2.5m telescope with a survey area of 14555 square degrees. BOSS spectrographs were used for the spectroscopic survey to measure the redshifts of 1.5 million luminous red galaxies and Lyman-alpha absorption towards 160,000 high-z quasars. 
The galaxy data comes from the Data Release 12 (DR12; \citealt{alam2015}). The data has been divided into two redshift ranges, LOWZ, which covers $0.0<z<0.4$, and CMASS, which is in the interval $0.4<z<0.7$. LOWZ and CMASS samples are further divided into LOWZ/CMASS North and LOWZ/CMASS South. The catalogs provide data on the coordinates, spectroscopic redshift, different fluxes, as well as several other parameters (\citealt{reid2016}).  

\subsection{Quasar Catalog}
We use the SDSS-DR 12 quasar catalog in order to have the same coverage for galaxies and AGN. The catalog is described in its entirety in \citet{paris2017} and contains the object's coordinates, redshifts obtained using different methods, the photometric SDSS magnitudes in u-, g-, r-, i-, and z-bands and their associated errors, and the absolute i-band magnitude. The spectroscopic redshifts are crucial to accurately measure the quasar-void distances. AGN selection methods relying on photometric properties, such as MIR, and matching with galaxy positions do not work well for our study of underdense regions because of the large distance uncertainties and small sample statistics.

\subsection{The Void Sample}
The voids used for this study are from the catalog of cosmic voids based on SDSS III Data Release 12, described in \citet{mao2017}. The voids are based on galaxies from the large-scale structure (LSS) catalogs, which are a part of the BOSS database. \citet{mao2017} use ZOBOV (\citealt{neyrinck2008}), which is an algorithm based on Voronoi tessellations and the watershed method (\citealt{platen2007}) to detect voids. Since voids are irregular in shape, each void has been assigned a center and an effective Voronoi radius which is the equivalent radius it would have if the volume of the underdense region was spherically symmetric. 

\begin{figure}
\includegraphics[width=10cm, height=8cm]{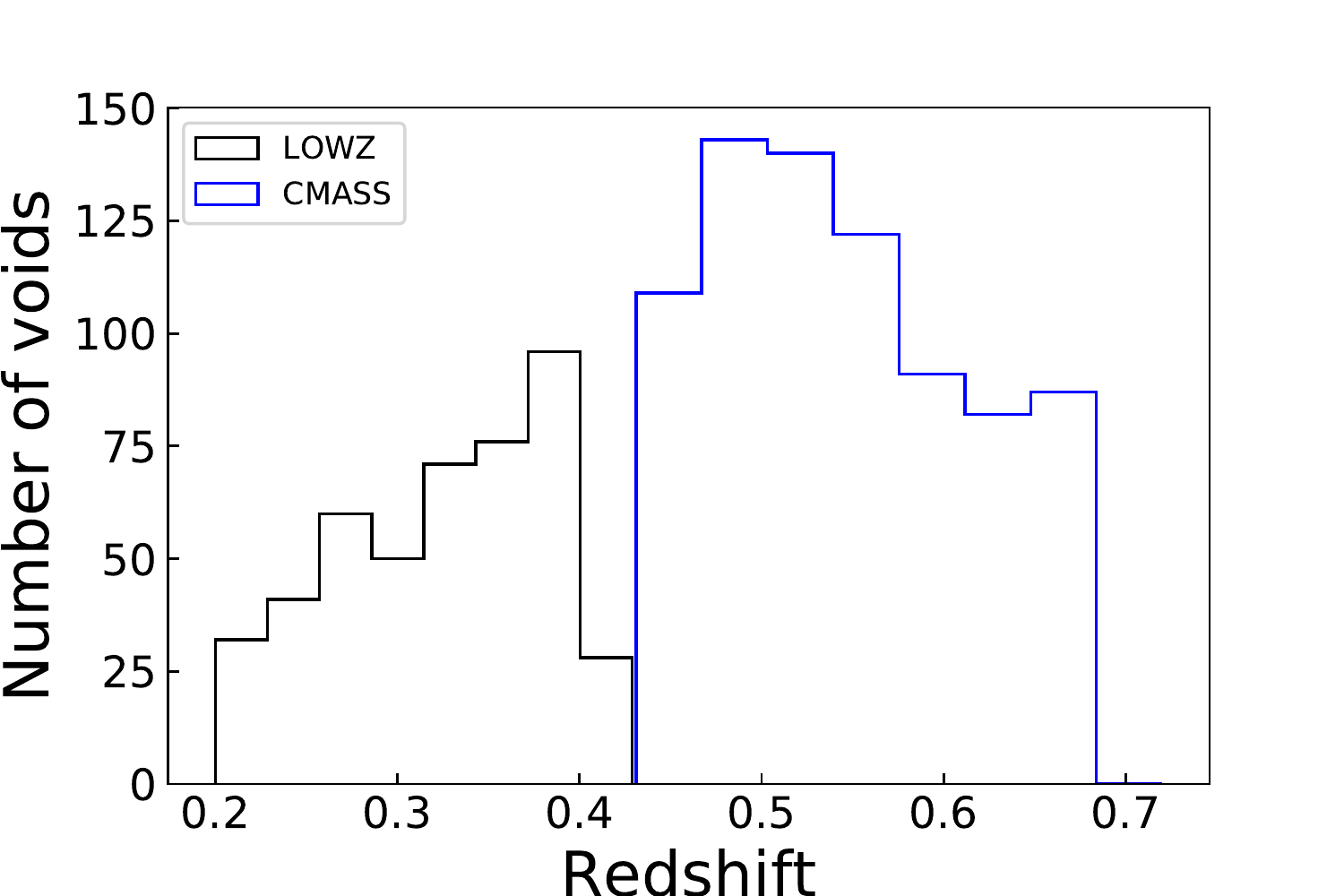}
\centering
\caption{Redshift distribution for the voids used in this study. \label{fig:hist}}
\end{figure}

The cosmic voids catalog from \citet{mao2017} provides the redshift of the weighted center of the void, the number of galaxies in the void, the total Voronoi volume, the effective void radius, the number density of the minimum density Voronoi cell in the void, the density contrast of the minimum density cell comparing to the mean density at that redshift, the ratio of the minimum density particle on a ridge to the minimum density particle of the void, the probability of the void origin as Poisson fluctuations, and the distance between the weighted void center and the nearest survey boundary \citep{mao2017}. 
The LOWZ voids are in the range $0.2<z<0.43$, whereas the CMASS voids cover the range $0.43<z<0.7$. The majority of the voids have effective radii ranging from 30--80 $h^{-1}$ Mpc (see Figure 2 in \cite{mao2017} for the void size distribution), where the LOWZ sample exhibits an extra tail distribution on the larger size compared to CMASS. Figure \ref{fig:hist} shows the redshift distribution for the cosmic voids sample used in this study for the LOWZ (low redshift) and CMASS (high redshift) samples. 

\subsection{Methodology}
In this study, the size of the voids was characterized by the effective radius, \reff{}, which 
lies in the range 30--80 $h^{-1}$ Mpc. 
Although individual voids can deviate from a sphere, \citealt{mao2017} found that stacking analyses using \reff\ provide stable and average void properties, such as the galaxy density. 
The voids are divided into five redshift intervals: $0.21-0.28$, $0.28-0.35$, $0.35-0.42$, $0.42-0.56$, and $0.56-0.70$, to study any evolutionary trend with redshift, if present.
Member galaxies and AGN for the voids were selected from the SDSS CMASS and LOWZ galaxy catalogs and the quasar catalog. 
To select void members, physical distances in the three-dimensional space are calculated between the void center and the galaxy center. We utilize the spectroscopic redshifts available for the voids and galaxies in the BOSS catalogs for more accurate distances. 
To classify the galaxies and AGN as members, we use a simple method based on the proximity of a galaxy to its nearest void center given by the physical distance between the galaxy and the void center. 
To investigate the differences in physical conditions in the inner and outer regions of the voids and see if environmental factors play a role in AGN abundance closer to the void center versus farther away, we assign each member galaxy to either the inner or outer region of the void. The criterion chosen for the inner and outer regions is based on the distance of a galaxy to the void center. The inner and outer void regions are defined to lie between $0.0-0.6$\reff\ and $0.6-1.3$\reff\ respectively, and their sizes vary with the effective radius of a void. 
This approach of assigning void membership to galaxies and defining inner and outer regions does not take into account the shape of the individual voids; however, for the purpose of a statistical study, the voids in each redshift bin are stacked and the effective shape of the resultant void is a sphere. 
For this study, we make an absolute magnitude cut in each redshift bin to select luminous galaxies, where the absolute magnitudes are calculated using the K-correction templates of \cite{assef2010}. The range used for the magnitude selection is absolute-r magnitude = $-23$ to $-22$ to focus on the luminous galaxies and is chosen to avoid the selection bias toward low luminosity galaxies at low redshifts. This provides the maximum number of galaxies for a statistical study of bright AGN and galaxies for all redshift bins. 
The fraction of AGN (or quasar fraction specific to this analysis), \fa, is defined as the ratio of the number of active galaxies and the total number of galaxies in the void.

\section{Results} \label{sec:sec3}
\begin{figure}
\includegraphics[width=10cm, height=10cm]{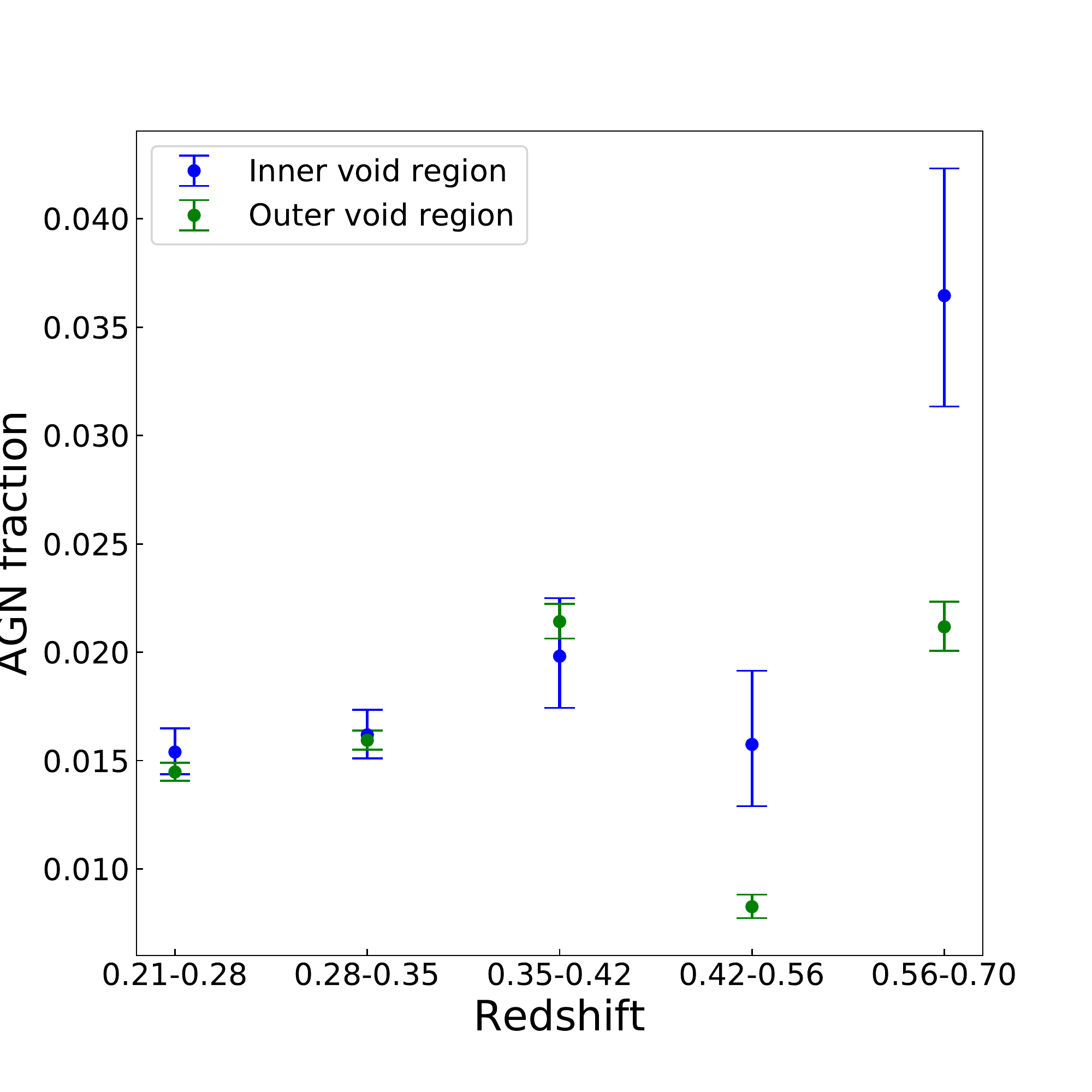}
\centering
\caption{AGN fraction in the inner and outer regions of the void for the five redshift intervals. The errors are Poisson errors that correspond to 1$\sigma$ confidence limits. \label{fig:fa_z}}
\end{figure}

We separate the voids in five redshift bins and calculate the AGN fractions in the inner and outer void regions in each redshift bin by stacking the voids in the bin. 
The AGN fractions for the inner and outer void regions for each redshift range are listed in Table \ref{tab:table1}, which also  provides the raw galaxy counts for AGN and non-AGN in the voids. The uncertainties associated with \fa\ are Poisson errors with 1$\sigma$ confidence limit. Figure \ref{fig:fa_z} shows the fraction of AGN in inner and outer void regions for each redshift bin. 
Our goal is to independently investigate the low redshift range (the LOWZ sample) and higher redshift range (the CMASS sample). At low redshifts, $z \leq 0.42$, we do not see a significant difference in the AGN abundance in the inner and outer regions of the void. However, for the two highest redshift bins, $0.42<z<0.56$ and $0.56<z<0.72$, the AGN fraction in the inner void region is significantly higher than for the outer void region. The differences in the AGN fractions between the inner and outer regions for the highest two z bins are $0.0152\pm0.0049$ and $0.0192\pm0.0053$, at $3.1\sigma$ and $3.6\sigma$ significances, respectively. Considering that we have five independent tests for the five redshift bins, the significance of the deviation between \fa\ values are $2.6\sigma$ and $3.1\sigma$ sigma, respectively for these two redshift bins.
The average AGN fraction across all redshifts for the inner void region is $0.028\pm0.004$, which is 1.3 times higher than that for the outer void region, where the average fraction is $0.021\pm0.001$.

We further divide the inner and outer regions into smaller bins to investigate the dependence of AGN fraction on the distance from the center of the cosmic voids. The distances are measured in \reff{} and range from the void center to 1.3\reff. 
The size of the radial bins is fixed at 0.1\reff\ except for the first bin, which is 0.3\reff. This was done to increase the signal-to-noise ratios as there are very few galaxies near the center of the void. The greater size of the first radial bin does not affect our analysis results since the motivation is to investigate any difference in AGN abundance that might be present between the inner and outer void regions and cosmic voids have very large effective radii. 

\begin{figure}
\includegraphics[width=12cm, height=22cm]{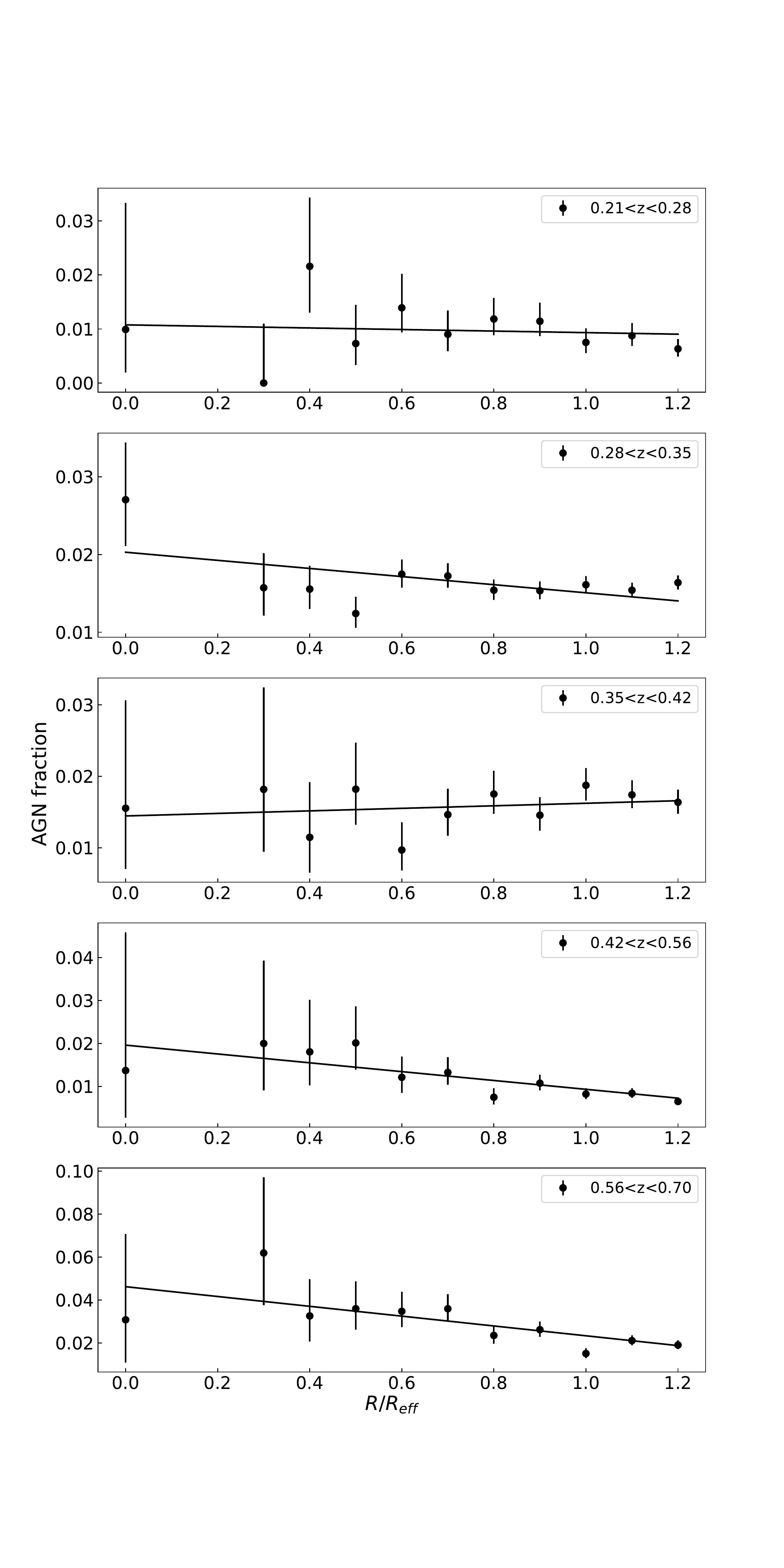}
\centering
\vspace{-0.8cm}
\caption{AGN fraction as a function of the distance from the center of the void in the units of \reff\ (the effective radius of the void region) for the five redshift intervals. A best-fit linear model (solid black line) has been calculated for each bin. \label{fig:radial}}
\end{figure}

Figure \ref{fig:radial} shows the AGN fraction plotted as a function of distance of the member galaxies from the void center for the five redshift intervals. 
The slopes and $1 \sigma$ uncertainties, where the amplitude uncertainty has been marginalized, for the best-fit lines are $-0.001 \pm 0.004$, $-0.005\pm 0.002$, $0.004 \pm 0.003$, $-0.010 \pm 0.006$, and $-0.023 \pm 0.008$, respectively, for the five redshift bins $0.21-0.28$, $0.28-0.35$, $0.35-0.42$, $0.42-0.56$, and $0.56-0.70$.
For the first three low redshift bins, we see some scattered results for $f_A$ from the center of the void out to 1.3\reff. The first and second redshift bins show a slightly negative trend, especially for the second bin where the significance is $\sim2\sigma$. The third bin has a positive slope which is not significant compared to measurement uncertainties. 
For the two higher redshift bins, we see $\sim 2\sigma$ decreases in AGN fractions with distance from the void center by comparing the measured slopes and uncertainties.

\begin{deluxetable}{lcrccc}[b!]
\tablecaption{AGN fraction in inner and outer void regions for the five redshift bins \label{tab:table1}}
\tablecolumns{6}
\tablenum{1}
\tablewidth{0pt}
\tablehead{
\colhead{Redshift\tablenotemark{a}} & 
\colhead{Inner void} &
\colhead{Outer void} & 
\colhead{Slope\tablenotemark{b}} &
\colhead{AGN\tablenotemark{*}} &
\colhead{Non-AGN\tablenotemark{*}} \\
}
\startdata
0.21 $\leq$ $z$ $<$ 0.28 & $0.012_{-0.003}^{+0.003}$ & $0.0086_{-0.0008}^{+0.0009}$ & $-0.001_{-0.004}^{+0.004}$ & 19;94 & 1585;10820 \\
0.28 $\leq$ $z$ $<$ 0.35 & $0.016_{-0.001}^{+0.001}$  & $0.0159_{-0.0004}^{+0.0004}$ & $-0.005_{-0.002}^{+0.002}$ & 219;1319 & 13309;81436 \\
0.35 $\leq$ z $<$ 0.42 & $0.013_{-0.002}^{+0.003}$ & $0.0167_{-0.0008}^{+0.0009}$ & $0.004_{-0.003}^{+0.003}$ & 36;372 & 2660;21807 \\
0.42 $\leq$ z $<$ 0.56 & $0.035_{-0.005}^{+0.006}$ & $0.0198_{-0.0010}^{+0.0011}$ & $-0.010_{-0.006}^{+0.006}$ & 39;368 & 1073;18132 \\
0.56 $\leq$ z $<$ 0.70 & $0.062_{-0.006}^{+0.007}$ & $0.0428_{-0.0017}^{+0.0017}$ & $-0.023_{-0.008}^{+0.008}$ & 77;627 & 1163;14021 \\
\enddata
\tablenotetext{a}{All the data in each $z$ bin is co-added.}
\tablenotetext{b}{The slope of AGN fraction as a function of void-centric distance from inner to outer regions.}
\tablenotetext{*}{Inner region; outer region}
\end{deluxetable}


\vspace{-0.25cm}

\section{Discussion and Conclusions} \label{sec:sec4}
We found a significantly higher AGN fraction in the inner void regions in the two higher redshift bins \textbf{($0.42<z<0.72$)} either using small bin sizes or separating the voids into two regions.  
There is a 50\% increase in \fa\ in the inner void region for the higher redshift voids, and we observe a decrease in AGN abundance as we move away from the void center.
For the lower redshifts, the second bin ($0.28<z<0.35$) shows a moderate decreasing trend ($\sim 2\sigma$) in Figure \ref{fig:radial}, whereas the first and third $z$ bins do not show a significant dependence on voidcentric distance when the uncertainties are taken into account. 
Although the difference in the analysis conclusions in the five redshift bins coincides with the LOWZ and CMASS boundary, where we measured environmental dependent AGN fractions in CMASS voids but largely no dependence for LOWZ counterparts, we did not identify potential biases affecting our analysis results between CMASS and LOWZ samples, since our analysis was focusing comparing different void regions within the same redshift bin and potential biases in galaxy, AGN, and void selections will cancel out in the comparison.
There are differences in selection of galaxies for LOWZ and CMASS samples, where the LOWZ galaxy sample selects bright, red galaxies and CMASS, similarly, targets luminous, red galaxies, along with extending to some blue galaxies. While this limits our analysis to more luminous red galaxies for both low and high-z ranges, the selection bias is consistent for inner and outer void regions. 
The void size distributions between LOWZ and CMASS samples are slightly different; however, again, since we are comparing the AGN fractions in the inner and outer void regions, it is unlikely these size difference can contribute significantly to our analysis results.
The redshift range $z<0.7$ of this study is relatively narrow compared to the parent quasar catalog with $z \le 7$, and the quasar selection at $z<0.7$ mainly relies on the UV excess approximated by the $u-g$ color (\citealt{ross2012}).  Thus, it is unlikely that the biases from quasar selections at $z<0.7$ contribute to the differences in our analysis between LOWZ and CMASS samples. 
Thus, the weaker environmental dependence for the three lower redshift bins could be physical, where the void environment close to the void center and in the outer parts responsible for AGN triggering has become more homogeneous as voids evolved, or it could be a result of the sample size. The spectroscopic SDSS data, while it provided more accurate redshifts, limited the sample to a smaller size, preventing us from measuring a minute trend smaller than the one observed in the two high redshift bins.

Our result of finding an increasing AGN fraction as we move toward the center of the void, in some redshift bins, provides a baseline result to be tested further with future studies. 
In the overdense regions, it is a better established result that the AGN abundance is lower in clusters and groups compared to the surrounding field regions at $z<1$ \citep[e.g.][]{2017MNRAS.472..409L,mishra2020}. Combining the result from this analysis, where we found hints of higher AGN fractions at void centers, we propose a thesis that at low redshifts where the AGN activities have declined from the cosmic peak, there is an overall trend of decreasing AGN activities with increasing environment density from the lowest void to the highest cluster regions, and this can be further confronted with future analysis using larger data sample. 

Study conducted by \citet{constantin2008} showed no change in AGN abundance from voids to walls for massive, luminous galaxies (\Mr  $\lesssim -20$ and log M/\msun $ > 10.5$) at very low redshifts which is in agreement with our results of finding no significant trend for lower redshifts. For moderately massive and luminous galaxies (\Mr  $\sim -20$ and log M/\msun $ < 10.5$), \citet{constantin2008} found higer AGN abundance in voids than walls, especially for low-luminosity AGN.

According to the model that the inflow of cold gas toward the center of the galaxy and accretion onto the central engine regulates black hole growth \citep[e.g.][]{2005Natur.433..604D, 2005ApJ...625L..71H, zanisi2021}, the availability of gas in these systems would determine the AGN activity.
Since we have selected the luminous galaxies (\Mr$<-20$) in both inner and outer void regions, they have more matter available to feed the central AGN. It is possible that the galaxies in the inner void region are able to hold onto their gas reservoirs in the absence of less gravitational disruptions resulting from the interactions with nearby systems as would be the case for outer regions of the void, where galaxies near and in the walls are not as isolated. This could potential explain the higher AGN fractions in the void centers observed at the two higher redshift bins.

It is worth noting that our study is based on the BOSS CMASS catalog to select galaxies at high redshifts which selects more red galaxies. This may result in the selection of more galaxies in the outer regions of the void since galaxies that are found near the void center tend to be bluer and faint \citep[e.g.][]{bruton2020,rojas2004}. However, this selection bias would affect the AGN and normal galaxies equally and thus, would not skew the \fa toward a higher value for the outer regions. 
We also measure the trend of AGN abundance relative to blue galaxies with voidcentric distance for the two highest redshift bins for discussion purposes. For this, we used the results from \citet{bruton2020} which is a comparative study of the relative abundance of red and blue galaxies in voids. \citet{bruton2020} uses the CMASS data and galaxies from the WiggleZ Dark Energy Survey \citep{drinkwater2010} to study the density profiles of the red and blue galaxies as a function of the void radius. They report lower red-to-blue galaxy ratio at void centers, and their relation can be used to convert the AGN fraction for CMASS galaxies investigated in this paper to the AGN fraction with respect to blue galaxies. After doing that, we do not find a significant trend for \fa\ relative to the blue galaxies with distance from the center of the void within error bars, in contrast to the decreasing AGN abundance from inner regions to the outer regions found in our study. This might be indicative of differences in the evolutionary track or environmental dependence of the red versus the blue galaxies in the voids or simply the case where signal-to-noise is low to make any conclusions. A larger sample of blue galaxies will be needed to study the AGN abundance in the blue galaxies.
The red-to-blue galaxy ratio in LOWZ voids are not reported in \citep{bruton2020}, which precludes us from estimating AGN-to-blue galaxy ratios in LOWZ voids.

While the literature for AGN activity and evolution of active galaxies in overdense regions, like clusters and groups, and their surrounding fields, is rich, the field of investigating the AGN evolution in cosmic voids remains greatly unexplored, but is extremely important to study the impact of local and global environmental factors that might play a role in triggering the nuclear activity in the central engine. The triggering mechanisms can also provide useful insight into the different conditions present in the cosmic voids and their evolution as compared to the high-density regions of the Large Scale Structure.
Future studies using larger samples of spectroscopic data will be able to increase the signal-to-noise ratio and constrain the void AGN abundance better at lower redshifts and enable the expansion of this analysis to a much larger redshift range. 

\acknowledgements
We are grateful to the anonymous referee for the helpful comments and recommendations that made the paper clearer and stronger. 
We acknowledge the financial support from the NASA ADAP programs NNX15AF04G, NNX17AF26G, NSF grant AST-1413056. 
\end{document}